\documentclass[square]{ws-procs11x85}
\usepackage{balance} 

\def\Journal#1#2#3#4{{#1} {\bf #2}, #3 (#4)}

\newcommand{\met}{\hbox{E\kern-0.5em\lower-0.1ex\hbox{/}}_T}
\newcommand{\psr}{PSR~B1259-63/SS2883}

\begin{document}

\twocolumn[
\title{Hydrodynamics of interaction of pulsar and stellar  winds and its impact on the high energy radiation 
of  binary pulsar systems}

\author{D.V. Khangulyan}
\address{Max-Planck Institute fur Kernphysik, Heidelberg, Germany\\E-mail: dmitry.khangulyan@mpi-hd.mpg.de}
\author{F.A. Aharonian}
\address{Dublin Institute for Advanced Studies, Dublin, Ireland,\\Max-Planck Institute fur Kernphysik, Heidelberg, Germany}
\author{S.V. Bogovalov}
\address{Moscow Engineering Physics Institute (state university), Moscow, Russia}
\author{A.V. Koldoba}
\address{Institute for Mathematical Modeling RAS, Moscow, Russia}
\author{G.V. Ustyugova}
\address{Keldysh Institute of Applied Mathematics RAS, Moscow, Russia}

\begin{abstract}
The hydrodynamics  of the interaction of pulsar and stellar winds in  binary systems harboring a pulsar 
and its impact  on the nonthermal  radiation  of the binary pulsar PSR~B1259-63/SS2883 is 
discussed.  The collision of an ultrarelativistic pulsar wind with a nonrelativistic
stellar outflow results in significant bulk acceleration of the shocked material from the
pulsar wind. Already at distances comparable to the size of the binary system, 
the  Lorentz factor of the shocked flow can be  as 
large  as $\gamma\sim4$. This results in  significant anisotropy of the inverse Compton 
radiation of accelerated electrons. Because of the Doppler boosting of the produced radiation,
one should expect a variable gamma-ray signal from the system.  In particular, this effect 
may naturally explain the reported tendency of a decrease of  TeV gamma-ray flux close to the periastron.
The modeling of the interaction of pulsar and stellar winds 
allows  self-consistent calculations of  adiabatic losses. Our results 
show that adiabatic losses dominate over  the radiative losses. These results  have  
direct impact  on the orbital variability of radio, X-ray and gamma-ray signals 
detected from the binary pulsar  \psr{}.
\end{abstract}
\vskip12pt  
]

\bodymatter

\section{Introduction}
Binary systems represent an important population of very high energy (VHE) 
sources. Presently this category comprises  four objects \cite{hess,ls5039, ls50392,lsi,cyg}. 
At least one of them,  PSR~B1259-63/SS2883, is a binary pulsar system consisting of 
a  $48\, \rm ms$ pulsar in  an elliptic orbit around a massive B2e optical star \cite{johnston1}. The 
multiwavelength observations of the system  
\cite{johnston,connors,johnston05,cominsky,kaspi,hirayama,chernyakova}
indicate correlations between different energy bands \cite{neronov}. One of the interesting  features
of the high energy  emission of  PSR~B1259-63/SS2883 is the  behavior of the gamma-ray light-curve which 
deviates from the earlier  theoretical predictions \cite{kirk99,kawachi04}.    The  
understanding of the nature/origin of the  high energy radiation of 
this object  is a quite complex issue.  It requires not only 
proper modeling  of the acceleration and  radiation process in the context 
of multiwavelength properties of the source, but also  
detailed studies  of the hydrodynamics of the interaction of the pulsar and stellar winds.  Indeed, 
the particle acceleration regime and formation of  energy distribution of relativistic particles after  
termination of the pulsar wind strongly depend on the dynamics  of the flow. 
In this regard one should note that the recent attempts at explanation  of the 
gamma-ray light-curve  of  PSR~B1259-63/SS2883 \cite{neronov,khangulyan07} 
contain assumptions  which are closely linked to the  hydrodynamics of the system. 
For example, 
the density changes along current lines determine  the adiabatic loss rate in the flow.  
The flow bulk  motion is also responsible for the particle advection (escape) from the system. 
The nonradiative losses caused by these processes may 
have a strong impact on the energy spectrum and  light-curve of gamma-rays \cite{khangulyan07}. 
Therefore, any self-consistent treatment of  the high energy radiation of the system 
requires hydrodynamic calculations of key parameters characterizing the interaction of 
two winds. This concerns, in particular,  the magnetic field. 
The magnetic field is expected to be nonhomogenous in the pulsar wind nebulae \cite{kc}. The calculations
of the spatial distribution of the strength of the magnetic field generally requires a  proper  
magnetohydrodynamical (MHD) approach.  However,  in the case when the 
the impact of the magnetic field on the  dynamics of  plasma is not dramatic,  it 
can be treated  hydrodynamically   within a  ``froze-in''  approach.  Below we assume the influence of the 
magnetic field to be weak and we discuss the impact of hydrodynamics  characterizing the collision of pulsar and stellar  winds
on the high energy radiation of the system.  
The results are based on the recent hydrodynamic  calculations 
of Bo\-go\-va\-lov et al. \cite{bogovalov08} conducted through  
two different approaches using  a relativistic code  in the ``pulsar zone''  and 
a nonrelativistic code  in  the ``optical star'' region. It is assumed 
that  initially both the pulsar and stellar winds expand radially. 

\section{Hydrodynamics of interaction of relativistic and nonrelativistic winds}

\begin{figure}[t]
\center
\centerline{\psfig{figure=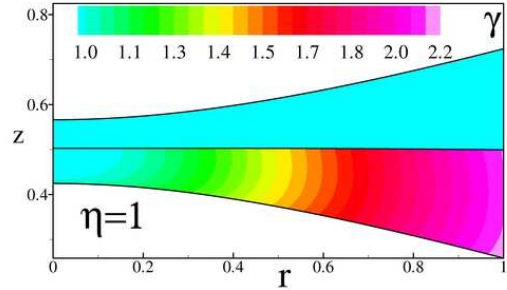,width=7truecm}}
\caption{The post shock flow for $\eta=1$. The color represents the bulk Lorentz factor. Parameters $r,z$ 
correspond to a cylindrical coordinate system [all lengths are measured in the units of the separation distance $D$; the pulsar and the optical star are located at $(r,z)=(0,0)$ 
and $(r,z)=(0,1)$, correspondingly].\label{eta1}}
\end{figure}

\begin{figure}[t]
\center
\centerline{\psfig{figure=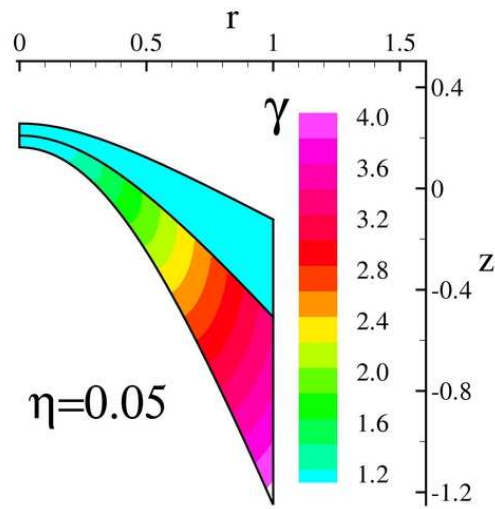,width=7truecm}}
\caption{The post shock flow for $\eta=0.05$. The color represents the bulk Lorentz factor. Parameters $r,z$ 
correspond to a cylindrical coordinate system [all lengths are measured in the units of the separation distance $D$; the pulsar and the optical star are located at $(r,z)=(0,0)$ 
and $(r,z)=(0,1)$, correspondingly].\label{eta2}}
\end{figure}

The discussion below concerns  the regions of linear scales comparable to the size of  the 
binary system (because  the variable synchrotron and IC radiation components are 
predominantly generated in this region). The results are based on the 
hydrodynamical calculations  reported by Bogovalov et al.   \cite{bogovalov08}.
The properties of the steady-state 
flow basically depend only on one parameter, namely on the
wind ram pressure ratio $\eta$ \cite{bogovalov08}:
\begin{equation}
  \eta={\dot E_{\rm sd}\over \dot Mcv_0} \ .
\label{eq_eta}
\end{equation}
Here  $\dot E_{\rm sd}$ is the pulsar spindown luminosity; $\dot M$ and $v_0$ are the stellar mass loss rate and the stellar outflow velocity, 
respectively (both winds are assumed to be isotropic). In the particular case of \psr{} the parameter 
$\eta$ is expected to be within  $10^{-2}$~--~$1$  \cite{bogovalov08}. 
Depending on the value of $\eta$, the structure of the flow may significantly change. It is demonstrated in  
Figs. \ref{eta1} and \ref{eta2}  for $\eta=1$ and $\eta=0.05$ (the color represents the flow bulk Lorentz factor).
Figures \ref{eta1} and \ref{eta2} show  that at distances comparable to the size of  the binary system, 
the  bulk Lorentz factor of the shocked flow  may  be as high as $\gamma\sim4$.  Because of Doppler boosting, 
this should result in a significant anisotropy of nonthermal emission of relativistic particles. 
Moreover, the  large  bulk
Lorentz factors  imply  significant  adiabatic losses. Indeed, the full energy per particle $\gamma w/n$
 is conserved along the flow line (here $w$ and $n$ are the  enthalpy and
particle densities).  Thus the acceleration of the flow is reduced to the transfer of the thermal energy
to the bulk motion, i.e. to adiabatic losses. Below we  compare the adiabatic loss time to other relevant 
time-scales, the particle escape time and IC  cooling time. 
Note that all these time-scales depend, unlike the overall structure 
of the flow,  on the separation distance  $D$ between stars. 

\begin{figure}[t]
\center
\centerline{\psfig{figure=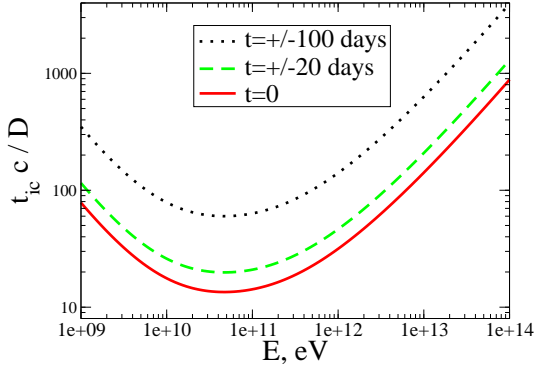,width=6truecm,angle=-90.0}}
\caption{Inverse Compton cooling time for three different epochs: 0, $\pm20$ and $\pm100$ days to periastron passage. 
$D$ is the  corresponding separation distance.\label{ic}}
\end{figure}

As long as the post-shock flow velocity is close to the speed of light $c$, the advection time scale can be estimated as
\begin{equation}
 t_{\rm esc}=\frac{D}{c}\,.
\end{equation}
It is convenient to express all  times in  the unit of $D/c$. In Figure \ref{ic} we show the 
IC cooling time for three different separations, which correspond to $\pm100$, $\pm20$ and $0$ days 
to the periastron passage. The calculations are performed for the following parameters: 
temperature of the optical star 
$T_*=2.2\times10^4$~K; eccentricity of the orbit $e=0.87$; periastron separation $D_0=9.6\times10^{12}$~cm.
As it follows from Fig.~\ref{ic}, the inverse Compton cooling time $t_{\rm ic}\gg D/c$, i.e. the radiative 
cooling time is much larger than the particle escape time, especially at large separation distances 
(obviously, $t_{\rm ic}\propto D^2$, due to decrease of target photon density).

Although it is not possible to obtain, from first principles,  the adiabatic loss rate, the $D$-dependence of the adiabatic
cooling time can  be defined. Interestingly, the adiabatic loss time has a  $D$-dependence similar to  
$ t_{\rm esc}$. Indeed, since adiabatic losses, which occur between two points on the  current line, depend only on the
ratio of densities in these points, one obtains 
\begin{equation}
  t_{\rm ad}\propto D\,,
\end{equation}
since the traveling time between these two points is proportional to $D$. This allows us to determine the $D$-dependence of
the ratio of IC to nonradiative cooling time:
\begin{equation}
 \frac{t_{\rm ic}}{t_{\rm esc/ad}}\propto D\,.
\label{ratio}
\end{equation}
Thus, in a case of collision of two isotropic winds small separation distances are more preferable for VHE gamma-ray production. 
Although, we note here that it does not necessarily imply higher observable fluxes, due to the anisotropy related to  
the Dopper boosting of the produced emission.

\begin{figure}[t]
\center
\centerline{\psfig{figure=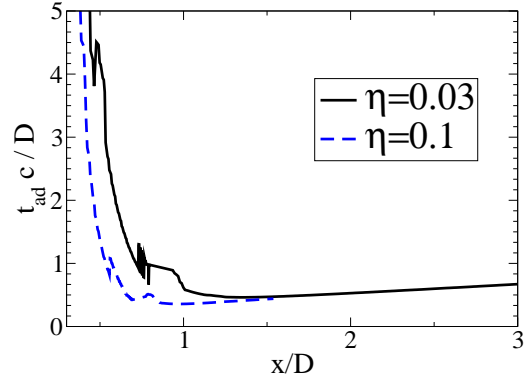,width=6truecm,angle=-90.0}}
\caption{Result of the numerical simulation of the adiabatic cooling time for two different values of $\eta$: 0.03 and 0.1. 
These calculations were performed along current lines which cross the pulsar wind termination shock close to 
the line connecting the stars. The coordinate $x$ corresponds to the distance along the current line.
\label{adiabat}}
\end{figure}

Generally, the adiabatic loss rate is not homogenous in the flow. In Figure~\ref{adiabat} we show typical dependences of
adiabatic cooling time, obtained with numerical simulation, along current lines for two cases: $\eta=0.03$ and $\eta=0.1$ 
(please note that sharp peaks/deeps in this figure are numerical artifacts). 
These calculations were performed along current lines close to the contact 
discontinuity in the ``pulsar zone''.
The  calculations show that for the flow structure  predicted by  the hydrodynamical  calculations, 
one should expect very fast (compared to  the IC cooling time) adiabatic losses over the entire orbit. 
This  should result in a strong suppression of nonthermal radiation of accelerated particles.

\section{Discussion}

The results presented above demonstrate the importance of hydrodynamics for the calculation of  
the high energy radiation of particles accelerated in the post-shock flow.    
Namely, 
(i) the adiabatic cooling time is significantly shorter than IC cooling time; (ii) the bulk 
Lorentz factor of the flow is rather large  ($\gamma\sim4$) already at distances, 
comparable to the size of the binary system. Both these effects have direct 
implication for  interpretation of observations of high energy
gamma- and X-ray emission components  from  the binary pulsar system \psr. 
In particular, the derived dependence in Eq.~(\ref{ratio}) (the ratio of IC cooling time to the nonradiative loss time
as a function of the separation distance) implies that 
the HD simulations do not support (at least in the particular case of two isotropic winds colliding) the 
assumption of  a sharp increase in the role of adiabatic losses 
towards  periastron as hypothesized in  ref.  \cite{khangulyan07} 
for explanation of the observed deep in the gamma-ray light-curve of \psr{} around periastron.
On the other hand,  the obtained absolute timescales of the adiabatic losses  (compared  
to the  Compton  cooling  times) show that the VHE signal should be significantly suppressed, especially 
at large separation distances.  The HESS observation of \psr{} \cite{hess} supports
this conclusion: the signal from the source was detected only during  epochs 
close to the periastron passage.

The effect of the flow bulk acceleration strongly modifies the  relationship between 
the synchrotron X-ray and inverse Compton gamma-ray fluxes produced  by the 
same population of relativistic electrons. Indeed, as long as the magnetic field is frozen 
in the plasma, its strength is determined by the structure of the flow and may significantly
vary across the flow. The plasma bulk motion affects the IC radiation production as well, 
but in a different way. Namely, in the co-moving frame, where electrons are distributed isotropically, the 
target radiation field will be Doppler boosted.  
This implies that hydrodynamical effects  can lead to a significant deviation from the standard relations 
between  the X-ray  and VHE gamma-ray fluxes (see e.g. \cite{atoyan,khangulyan2}).

Moreover, due to the large  bulk Lorentz factors (for details see \cite{bogovalov08}),
 the observed fluxes should have strong orbital phase dependence. Indeed, the 
direction of the post shock flow varies with the motion of the pulsar along 
the orbit around the  star. This implies significant changes of the Doppler factor
$\delta$.  Namely, $\delta \gg 1$ for small viewing angles (when the flow bulk velocity  
is directed towards the observer);  and $\delta \ll 1$ for large viewing angles.  Such a geometry 
is expected, in particular,  at the  periastron passage. Correspondingly, this will have a strong 
impact on the light-curve of nonthermal radiation of electrons, $F_\gamma \propto \delta^{n}$
where typically $n \geq 3$. Interestingly, in such a case one should expect  a direct
correlation between different energy bands (from radio to VHE), even though the electrons 
responsible for these energy intervals of electromagnetic radiation may not have the same origin. 
This conclusion is supported by 
multiwavelength observation of \psr{}\cite{neronov}.  Moreover, this effect could be the reason 
for the  deep in the  VHE light-curve   close to the periastron 
as reported by the HESS collaboration  \cite{hess}.

\balance

\end{document}